\pdfoutput=1
\documentclass[a4paper]{jpconf}
\usepackage{amsfonts,palatino,amsthm,latexsym,mathrsfs}
\usepackage[fleqn]{amsmath}
\usepackage{amssymb}
\usepackage{dsfont}
\usepackage{epsfig}
\usepackage{textcomp}
\usepackage{longtable} 
\usepackage{hyperref} 
\usepackage{multirow}
\usepackage{enumerate} 
\usepackage{color} 

\DeclareMathAlphabet{\mathpzc}{OT1}{pzc}{m}{it}
\def\eali\end{align*}
\def\be{\begin{equation}}
\def\ee{\end{equation}}
\def\bea{\begin{eqnarray}}
\def\eea{\end{eqnarray}}
\def\bal{\begin{align}}
\newcommand{\eal}{\end{align}}
\def\ble{\begin{flalign}}
\newcommand{\ele}{\end{flalign}}
\def\ba{\begin{array}}
\def\ea{\end{array}}
\def\bali{\begin{align*}}
\def\nn{\nonumber \\}

\def\bite{\begin{itemize}}

\def\eat{\end{itemize}}
\def\begfig{\begin{figure}[h!]}
\def\endfig{\end{figure}}

\newcommand{\bwe}{\begin{widetext} \begin{eqnarray}}
\newcommand{\ewe}{\end{eqnarray}\end{widetext}}

\def\tfrac#1#2{{\textstyle{#1\over #2}}}
\def\half{\tfrac{1}{2}}

\def\quarter{\tfrac{1}{4}}

\def\del{\partial}
\def\deli{\del_i}
\def\delj{\del_j}

\def\delt{\frac{\del}{\del t}}
\def\delx0{\frac{\del}{\del x^0}}
\def\delx1{\frac{\del}{\del x^1}}
\def\delx2{\frac{\del}{\del x^2}}
\def\delx3{\frac{\del}{\del x^3}}

\def\pdif#1#2{\frac{\del #1}{\del #2}}

\def\dag{\dagger}
\def\adag{a^{\dag}}
\def\dagg{{}^{\dag}}

\def\ket#1{\mid #1 {\cal{i}}}

\def\norm#1{\| #1 \|}

\def\vect{\overrightarrow}

\def\N{\ensuremath{\mathds{N}}}

\def\R{\ensuremath{\mathds{R}}}

\def\Lag{\ensuremath{\mathcal{L}}}

\def\half{\ensuremath{\frac{1}{2}}}
\def\so4{\ensuremath{\mathfrak{so(4)}}}
\def\su2{\ensuremath{\mathfrak{su(2)}}}
\def\SU2{\ensuremath{{SU(2)}}}

\def\gij{g_{ij}}

\def\rf#1{Eq.(\ref{#1})}

\def\ni{\noindent}

\def\th{\ensuremath{^{\text{th}}}}

\def\bd{\bf \begin{definition}: \it}
\def\ed{\end{definition} \rm}
\def\blem{\bf \begin{lemma}: \it}
\def\elem{\end{lemma} \rm}
\def\bthe{\bf \begin{theorem}: \it}
\def\ethe{\end{theorem} \rm}
\def\bcor{\bf \begin{corollary}: \it}
\def\ecor{\end{corollary} \rm}
\def\bpro{\bf \begin{proof}: \rm}
\def\epro{\end{proof} \rm}

\def\undback16{& \!\!\!\!\!\!\!\!\!\!\!\!\!\!\!\!}

\def\+{\ensuremath{\ket{+}}}
\def\-{\ensuremath{\ket{-}}}

\def\Hs{\ensuremath{\mathcal{H}}}

\begin{document}
\title{Local tuning of Coupling Constants allows for Quantum Fields in Curved Spacetime in the Lab}

\author{Isabeau Pr\'emont-Schwarz}
\ead{isabeau@aei.mpg.de}
\address{Max-Planck-Institut f\"ur Gravitationsphysik, Albert Einstein Institute,\\
Am M\"uhlenberg 1, Golm, D-14476 Golm, Germany}
\begin{abstract}
In this paper we will investigate how one can create emergent curved spacetimes by locally tuning the coupling constants of condensed matter systems. In the continuum limit we thus obtain continuous effective quantum fields living on curved spacetimes. In particular, using Stingnet condensates we can obtain effective electromagnetism. We will show for example how we obtain quantum electromagnetism ($U(1)$-Yang-Mills) in a black hole (Schwarzschild) spacetime.

\end{abstract}

\section{Introduction} \label{Intro}

One of the greatest mysteries of modern physics consists of how to deal with quantum mechanics and a curved dynamical spacetime at the same time. This is absolutely necessary in order to understand phenomena in which both gravitational and quantum effects are non-negligible, for example the early Big Bang or small black holes. Modern physics describes matter using the quantum field theory (QFT) we name the Standard Model (SM) and spacetime using Einstein's theory of general relativity (GR). The first is described mathematically in the language of linear algebra while the second is described mathematically in the language of differential geometry. For consistency, both theories need to work together since the curvature of spacetime is sourced by the matter fields and the quantum matter fields live on the spacetime. Making the two theories work together however is highly non-trivial because GR takes as input classical energy-momentum tensors, not quantum-operator-valued energy-momentum tensors and defining the Hilbert space in which the quantum fields live requires the knowledge of the entire spacetime \cite{QFTcurvedst} which cannot be known in GR unless the solutions to the matter fields are also know. There are many theories and models under the umbrella name of quantum gravity (QG) which are put forward as potential resolutions to the inconsistency of modern physics \cite{Strings,LQG,CDT,causalsets, spinfoams,GFT,Twistors, asymsafe} but none have passed any stringent experimental test, none are generally accepted.  

Fortunately or unfortunately (depending on one's perspective), the parts of the universe we have access to at the moment are extremely flat\footnote{Inflation in cosmology was proposed in part to solve the ``flateness problem''.}. Thus, because quantum effects are for the most part visible only at very small scales, when a quantum description of matter is necessary, we can usually neglect gravitation and assume that we live in a Minkowski spacetime. Inversely, when dealing with the large-scale, we neglect quantum effects and assume classical matter in GR. This however, means that it is difficult to investigate experimentally the effects of quantum fields in curved spacetime. An increasing amount of effort is being put into the
search for experimental evidence of quantum gravitational effects\cite{expqg} though no such evidence has
yet been found. Effects which are predicted to arise from having quantum fields in curved spacetime include Hawking radiation (the thermal radiation of black holes and other causal horizons)\cite{hawkrad} and cosmological particle creation (particle creation due to the expansion of the universe)\footnote{See \cite{CosmopartPrain}, for example, for an attempt to recreate this effect in the lab}. All of these effects have yet to be observed in Nature.  Of particular interest would be the observation of the predicted Hawking radiation . Even though it is only a kinematical effect of QG, it is one of the very few predictions of an essentially quantum-gravitational nature which is generally believed to be true. In order to overcome our lack of access to highly curved chunks of spacetime, analogue models have been proposed where a curved spacetime is simulated. 
In fact some experiments with analogue model black holes have already been made that claim to observe stimulated emission of Hawking radiation, though at a classical level \cite{Silkelab}.

The idea behind analogue models is simple and elegant. Take a physical system which you can work with in the lab such that in some limit, the equations of motions or the action for some of the degrees of freedom are the same as those of some form of matter on curved spacetime. The first model proposed was probably that of Gordon \cite{Gordonanalog} of flowing dielectrics which was later much refined\cite{Pham,Plebanskianalog,Felice}. Today, the most popular models are sound and surface waves in water\cite{Unruh81,Jacobsonsonicshort} for classical waves on a curved spacetime and Bose-Einstein condensates (BEC) \cite{Garay1, Garay2, BLV1} and superfluid Helium 3 ($^3$He) \cite{JacVolvHe,VolovikHe} for quantum fields on curved spacetime. Slow light \cite{LeonhardtSL}, a quantum remake of the earlier dielectric models using modern materials also seems to have a promising future. Analogue models are also a source of inspiration for fundemental research in quantum gravity where gravity is seen as emergent rather than fundamental (see for example \cite{trappedgraphity}).

Expanding on a previous conference proceeding\cite{emergproc}, what we would like to propose in this paper is a slightly different class of models which is obtained from a simple idea. Condensed matter systems can give rise to many types of quasiparticles, from Majorana fermions\cite{majoranareturns} to gauge bosons and anyons\cite{gaugebosqhall}, practically any type of matter one might want can be cooked up by some condensed matter system.  We will show in this paper that by varying the coupling constants, in space and in time\footnote{Space and time is, for the purpose of this sentence, understood to be the flat spacetime of the lab in which the condensed matter system is located.}, of a condensed matter system, we can simulate a curved spacetime for the quantum fields of the system. We can thus simulate any\footnote{Whereby ``any'' we mean any effective matter which can arise from a condensed matter system.} type of matter on almost any curved spacetime. 

In the next section we investigate how a variable speed of light can be reinterpreted as a curved geometry with a constant speed of light. In section \ref{Emergence} we show how an effective metric emerges from a tweaked quantum system. In section \ref{QEDBH} we suggest a concrete realization of U(1)-Yang-Mills theory in a Schwarzschild black hole background. 

%
%
%
%
%

\section{Variable Speed of Light or Curved Geometry: two descriptions of the same thing.}\label{Varspeed}
Let us suppose we have a D-dimensional space $\R^D$ which we endow with the flat metric $\delta_{i j}$. Let us also suppose that we have a field on the space which can propagate information at a finite speed $c(x;\overrightarrow{u})$ which is anisotropic and inhomogeneous in the sense that it depends on the position in space, $\overrightarrow{x}$, and the direction $\overrightarrow{u}$, $\overrightarrow{u}$ being a unit D-vector (with respect to the flat spacial metric). If this field is all that exists in the space, it may be difficult to justify the use of the flat metric $\delta_{\mu\nu}$ as it is not based on anything observable. A more operational (and arguably physical) definition of distance should rely on observables in the theory. A particularly natural choice in our case is to measure distances by sending signals with the field\footnote{In the case of multiple fields, what follows will only work if all the fields propagate information at the same speed.}, thus relating elapsed time with distance. In other words, if the speed of signalling at position $x$ and in direction $\vect{u}$ is $c(x;\vect{u})$, we redefine the distance along the path $y:[0,1] \rightarrow \R^3$ to be 
\begin{align}
 \int_0^1 \frac{\norm{\del y/\del s}}{c(y(s); (\del y/\del s)/\norm{\del y/\del s})}ds .\label{pathc}
\end{align}
Thus infinitesimal distances are obtained through the inverse speed and finite distances are obtained by integrating over geodesics.
By definition then, we may define the speed of signalling, which, by analogy, we will call the speed of light, to be $1$ at the expense of changing the metric. 
We thus started out with an flat spacetime metric
\begin{align}
 \eta_{i j} := -dt^2 + (dx^1)^2 + (dx^2)^2+(dx^3)^2 ,
\end{align}
but if distances are to be measured by sending signals, we are compelled to change that metric to 
\begin{align}
 g(x;v,v): = \varphi(x)^2 (-(v^0)^2 + {k^2(x,\overrightarrow{v}/\norm{\overrightarrow{v}})}\overrightarrow{v}^2 ), \label{newmetric}
\end{align}
where $x:=(x^0,\overrightarrow{x})$, is the spacetime position, $v:=(v^0,\overrightarrow{v})$ is a tangent vector at $x$ and $k(x,\overrightarrow{v}/\norm{\overrightarrow{v}}) := \frac{1}{{c(x;\overrightarrow{v}/\norm{\overrightarrow{v}})}}$ is the inverse of the speed at location $x$ and in direction $\overrightarrow{v}$. Notice how we have an arbitrary local conformal factor $\varphi(x)\in\R_+$ which we cannot fix classically since to impose that the speed of signalling is one, we have the option of rescaling space by $x\rightarrow 1/v x$ or rescaling time by $t\rightarrow v t$ or a mixture of both; however, as we will see, quantum mechanics give us a scale ($\hbar$) against which to measure and fix this local conformal degree of freedom. From \rf{newmetric}, given a position $x$ and two 4-vectors $v$ and $w$,  we naturally retrieve the inner product between $v$ and $w$ at $x$ as
\begin{align}
  g(x;v,w):= \half[g(x;v+w,v+w)-g(x;v,v)-g(x;w,w)]. \label{innerprod} 
\end{align}
This will be a tensor field - function of position $x$ - acting on two vectors if the variable speed satisfies the proper conditions. For \rf{innerprod} to define a metric, certain conditions on the speed $c$ need to be respected, namely linearity ($g(x;\lambda u,v)= \lambda g(x;u,v)$ and $g(x;u+v,w)=g(x;u,w)+g(x;v,w)$), since symmetry is trivially satisfied by the definition. 
Extending the function $k$ on the 2-sphere to $\R^3$ by defining
\begin{align}
\tilde{k}(\vect{v}):= k(\vec{v}/\norm{\vect{v}})\norm{\vect{v}}, 
\end{align}
linearity is equivalent to 
\begin{align}
 \tilde{k}^2(\vect{u}+ \lambda\vect{v})- \tilde{k}^2(\vect{u}) -\lambda^2\tilde{k}^2(\vect{v}) =  \lambda [\tilde{k}^2(\vect{u}+\vect{v})- \tilde{k}^2(\vect{u}) -\tilde{k}^2(\vect{v})] \label{scalin} \\
 \tilde{k}^2(\vect{u}+\vect{v}+\vect{w}) = \tilde{k}^2(\vect{u}+\vect{v}) + \tilde{k}^2(\vect{v}+ \vect{w})+ \tilde{k}^2(\vect{w}+ \vect{u}) -\tilde{k}^2(\vect{u})-\tilde{k}^2(\vect{v})- \tilde{k}^2(\vect{w}).\label{adlin}
\end{align}
Because we assume the speed $c$ to be finite, the metric \rf{innerprod}, if it exists, is non-degenerate.  Indeed many speed functions $k$ do not give rise to metrics (cf. Appendix \ref{speedmet} where we explicit the type of speed anisotropy compatible with \rf{adlin}.), but as we will see, generic Lagrangian derived systems do indeed give rise to varying speeds which can be reinterpreted metrically. 

Inversely, it is easy to see that given a curved spacetime and a foliation, one is always at liberty to reinterpret it as a variable speed of light $c(x,\vect{u})=\sqrt{-\frac{g(x;N(x),N(x))}{g(x;\vect(u),\vect(u)}}$ and a local conformal factor $-g(x;N(x),N(x))$ where $N(x)$ is the normal to the time slice at $x$. 

For now, let us simply say that subject to the proper constraints of the anisotropy on the varying speed of signalling, we may interpret variable speed as corresponding to signals going at a constant speed but seeing a curved spacetime which is determined up to a local conformal factor. In the next section we will see how we fix that conformal factor from the size of quantum fluctuations. 

\section{Emergence of Spacetime}\label{Emergence}
The recipe to for cooking up matter fields on curved spacetimes in the lab is conceptually rather simple. The first step consists of choosing the type of matter wanted. Secondly, one finds a condensed matter system which whose collective degrees of freedom reproduce the type of matter sought after; fermionic fields, scalar fields, and the like all have condensed matter systems from which they can emerge. As explicitly shown in \cite{LRBint}, the speed of signals can be varied by varying the coupling constants. How one locally varies the coupling constants depends on the exact system; as we will see later, in the example we give, we locally vary the coupling constants by changing the density and width of conductive sheets as well as the size of the holes in which the spins are placed. Thus the third step is to upgrade the coupling constants to functions of space and time, thereby making speed a function of spacetime. These functions can then be chosen so as to give the desired spacetime. Here is how it works in more detail. 

Let us imagine that we are given a quantum mechanical condensed matter system in D spacial dimensions.   In the limit where the lattice is very fine compared to measurements,i.e. the continuum limit, the discrete degrees of freedom can be approximated by continuous fields. Generically, the effective Lagrangian for the system will be of the form
\begin{align}
\Lag(x) = -\left(\sum_a \half G_a^{ij} \deli \phi_a(x) \delj \phi_a(x) + \half M_a^2\phi_a^2(x)\right) -V(\phi_a(x))\label{genlag}
\end{align}
if in the continuum limit the degrees of freedom correspond to scalar fields (for example in the case of coupled quantum oscillators on a lattice).  In what follows we will assume that the continuum (or emergent) degrees of freedom are scalar fields, but the reasoning is the same for other types of fields. In fact, for the concrete example we give in section \ref{QEDBH} we have a spin-1 $U(1)$ gauge-field (``light'') describing the continuum limit of the degrees of freedom. In \rf{genlag} repeated indices are summed over as per Einstein's summation convention. The indices $i$ and $j$ run from $0$ to $D$. $\del_0$ corresponds to the time derivative $\delt$ where $t$ is the time as measured by clocks in the Lab; $\deli$ with $i>0$ corresponds to a spacial derivative in one of the $D$ directions of space. $V$ is an function of the fields $\phi_a$ which is bounded from below (otherwise the resulting theory would be unstable and ill-defined). 

Without loss of generality we may suppose $G_a$ to be symmetric ($G_a^{ij}=G_a^{ji}$) since its antisymmetric part vanishes in any case in \rf{genlag} (for commutative momentum-spaces). The canonically conjugate momentum to the field $\phi_a(x)$ is then
\begin{align}
 \pi^a(x) = -G_a^{0 j}\delj \phi_a(x) . \label{conjmoma}
\end{align}
Thus the Hamiltonian is 
\begin{align}
 \Hs = & \sum_a \pi^a\del_0\phi_a - \Lag \nn
= & \sum_a \left[\begin{array}{cc} \del_0\phi_a & \vect{\del}\phi_a \end{array}\right] \left[\begin{array}{cc} -G_a^{0 0} & \vect{0} \\ \vect{0} & G^s_a \end{array}\right] \left[\begin{array}{c} \del_0\phi_a \\ \vect{\del}\phi_a \end{array}\right] + \sum_a \half M_a^2\phi_a^2(x) + V(\phi_a(x)) , \label{Habibi}
\end{align}
where $G^s_a$ is the spacial part of $G_a$ (i.e. $G_a$ with the zeroth line and column removed), $\vect{0}$ is the D-dimensional null vector, and $\vect{\del}$ is the D-dimensional spacial gradient. For the theory to be well defined, the Hamiltonian must be bounded from below, this implies that the matrices in \rf{Habibi} must be positive definite which in turns implies that $G_a^{0 0}<0$ and $G^s_a$ must be positive definite which means that the matrix $G_a$ must have Minkowskian, $(-,+, \ldots, +)$, signature.

If we now fine-tune our system so that the different fields $\phi_a$ all propagate at the same speed, that is all the matrices $G_a$ are equal, we may define
\begin{align}
 g^{ij} = \frac{G_a^{ij}}{(\det (-G_a))^\frac{1}{D-1}} \label{ginv} . 
\end{align}
If we additionally define $g_{ij}$ as the inverse of $g^{ij}$, $g= \det(-g_{ij})$, $m_a= \frac{M_a}{g^\quarter}$ and we redefine the potential $V(\phi_a)\rightarrow \frac{1}{\sqrt{g}}V(\phi_a)$ we may rewrite the Lagrangian of \rf{genlag} as
\begin{align}
 \Lag(x) = -\sqrt{g}\left\{ \left(\sum_a \half g^{ij} \deli \phi_a(x) \delj \phi_a(x) + \half m_a^2\phi_a^2(x)\right) -V(\phi_a(x)) \right\}. \label{newlag}
\end{align}
Not that since all the $G_a$ had Minkowskian signature, $\gij$ will also have the same Minkowskian signature. Written in this way, the Lagrangian looks a lot like the Lagrangian of fields in a curved spacetime with metric $g_{ij}$. The only difference being that here $\gij$ is a constant independent of $x$. However, $\gij$ being a collection of coupling constants, nothing stops of from tuning those coupling constants locally to make them depend on $x$, the spacetime location. If we do that we obtain a new Lagrangian
\begin{align}
 \Lag(x) = -\sqrt{g(x)}\left\{ \left(\sum_a \half g^{ij}(x) \deli \phi_a(x) \delj \phi_a(x) + \half m_a^2\phi_a^2(x)\right) -V(\phi_a(x)) \right\}. \label{newworld}
\end{align}
which this time is identical to the Lagrangian of fields in a curved spacetime defined by the Minkowskian metric $\gij(x)$. 

Now we see how the conformal factor of the metric is set by the size of quantum fluctuations. If we multiply the metric by a conformal factor $\Phi^2 $ (i.e. $\gij\rightarrow \Phi^2 \gij$) then the terms in the Lagrangian will be multiplied by $\Phi^{2D}$ or $\Phi^{2(D+1)}$. Since, in quantum physics the amplitudes are given by a sum of $e^{\frac{1}{\hbar}\int d^{D+1}\Lag}$, if the Lagrangian gains a factor of $\Phi^{2D}$ it is equivalent to having $\hbar\rightarrow \frac{\hbar}{\Phi^{2D}}$. Thus, for example, in 3 spacial dimensions, multiplying the effective time and space distances by 2 is equivalent to reducing the quantum fluctuations by a factor of $\frac{1}{64}$ \footnote{And in addition doubling the masses and quadrupling the potential term.}. Hence the conformal factor which remained undetermined classically is determined quantum-mechanically via the size of the quantum fluctuations. 

Hence we observe that locally modifying the coupling constants of matter fields in flat spacetime results in having does fields live on an effective curved spacetime manifold. Notice that we could also have allowed for the mass of the fields to depend on the spacetime location. One interesting point to note however is that the one thing which really have no choice about is the signature of the metric. If we start out with a Hamiltonian which is bounded from below\footnote{We talk here of the Bosonic case, the Fermionic case is of course more complicated.}, then the signature of the effective metric must be Minkowskian. This is very intriguing and might be telling us about some deeper more intricate relation between quantum physics and relativity. This, especially considering that, as previously mentioned, it is the size of quantum fluctuations which determine the conformal factor of the metric. In the following section we give a concrete realization of U(1) Yang-Mills on a Schwarzschild black hole background. 

\section{QED in Schwarzschild Spacetime}\label{QEDBH}
In \cite{WenLab} Wen proposed a concrete realization of his string-net condensate model for U(1) gauge Yang-Mills.  In this section we will show how one can take this in-lab condensed matter system of emergent light and locally tune the coupling constants of the system in order to end up with emergent light on an effective Schwarzschild black hole background. Thus creating a quantum system of a black hole with electromagnetic radiation. In addition to the interest of being able to observe a purely quantum gravitational\footnote{Even if only kinematically gravitational.} effect, because the underlying quantum system will consist of a spin lattice, we will be implementing a minimum length scale. That is interesting because there has been some questioning \cite{transplanckjac} as to whether Hawking radiation would exist if there was a minimal length scale due to the fundamental way in which the continuum plays a roll in deriving the Hawking radiation\footnote{One requires the existence of trans-Planckian modes in the original derivation.}.  
 
The conceptual idea behind string-net condensate models is as follows (see \cite{WenBook,WenLab,LRB1} for technical details). The underlying quantum system is a spin lattice or quantum rotor lattice. That is, a lattice, with on each edge, a quantum rotor or a spin. A string operator corresponding to a path $\gamma$ on the lattice corresponds to the ordered product of alternating raising and lowering operators along the path. So, for instance, in the case where we have have spin $J$s on the edges of the lattice then the string operator can be written:
\begin{align}
 S_{\gamma} = a_{\gamma_1} \adag_{\gamma_2}a_{\gamma_3}\ldots a^{\half(1+ (-1)^{n})(\dag) + \half(1+ (-1)^{n+1})}_{\gamma_n} ,
\end{align}
where $\gamma_j$ labels the $j\th$ edge of the path and 
\begin{align}
 a_k \propto & S_k^x + i S_k^y \nn
\end{align}
with $\{S_k^x,S_k^y,S_k^z\}$ being a basis of the lie algebra \su2\ acting on the Hilbert space of the spin on lattice edge $k$ such that $[S_k^x, S_k^y]= S_k^z$ (and cyclic permutations) is satisfied. The typical string-net condensate Hamiltonian then consists of three terms: a string tension term ($\propto \sum_k (S_k^z)^2$ in our spin $J$ example), a string fluctuation term ($\propto \sum_{\gamma} S_{\gamma}+S_{\gamma}^{\dag}$, the sum of all string of length 2 operators), and a Gau\ss{} constraint term ($\propto \sum_v \left(\sum_{e \ni v} S_e^z\right)^2$, in our example, where $v$ is a vertex of the lattice and $e\ni v$ is an edge which is attached to that vertex) which energetically penalizes the ends of open strings. When the coupling constant in front of the Gau\ss{} constraint is much bigger than the other two coupling constants, we effectively project down on to the sub-Hilbert space where the constraint is imposed exactly and thus there are no open strings. In particular, the the string fluctuation term for the effective Hamiltonian becomes a sum over plaquettes $P$ of the closed-string operators around the plaquettes\cite{WenLab}. In other words 
\begin{align}
 H_{\text{eff}} = J  \sum_e (S_e^z)^2 - \half g \sum_P (W_P + W_P\dagg) ,\label{Heff}
\end{align}
where the first sum is over the edges of the lattice and $W_P$ is the closed-string operator over plaquette $P$. If, on the edges of the lattice, we have quantum rotors $\theta_e$\footnote{At low enough temperatures, spins approximate quantum rotors.} then $S_e^z = -i\pdif{}{\theta_e}$, $a_e = S e^{-i\theta_e}$ ($S\in\half\N$)and the continuum limit of the effective theory, when $J\ll g$, is $U(1)$-Yang-Mills, in $2+1$ dimensions with continuum Lagrangian
\begin{align}
\Lag_{2+1} =  \frac{\vect{E}^2}{J} - l^2 g B^2 ,\label{Lconsimple} 
\end{align}
where $l$ is the size of the lattice spacing and the coupling constants are given modulo constants of order one which depend on the exact lattice configuration. This Lagrangian is arrived at by defining the gauge field with support on lattice edges $\vect{A_e} := \theta_e \vect{e}$, where $\vect{e}$ is the unit vector pointing in the direction of edge $e$ and then taking the continuum limit. We can re-write \rf{Lconsimple} as 
\begin{align}
 \Lag_{2+1} = -\sqrt{\det(g_{2D})} g_{2D}^{ik} g_{2D}^{jm}F_{ij}(A)F_{km} , \label{L2D}
\end{align}
where the metric is (in Cartesian $\{t,x,y\}$ coordinates)
\begin{align}
 g_{2D}:= \left[\begin{array}{ccc}
                -J^2 & 0 & 0 \\
		0 & \frac{J}{g l^2} & 0 \\
		0 & 0 & \frac{J}{g l^2}
               \end{array} \right] \label{g2Dxy}
\end{align}
or in polar coordinates $\{t,r,\phi\}$
\begin{align}
 g_{2D}= \left[\begin{array}{ccc}
                -J^2 & 0 & 0 \\
		0 & \frac{J}{g l^2} & 0 \\
		0 & 0 & \frac{J}{g l^2} r^2
               \end{array} \right] .\label{g2Dpol}
\end{align}

It is also possible to make a 3+1 dimensional system of emergent light \cite{WenBook,WenLab} by layering $2D$ lattices together so make a $3D$ lattice. In which case we obtain the following continuum Lagrangian, written in orthonormal coordinates, and with the third coordinate perpendicular to the layering
\begin{align}
\Lag_{3+1} =  \frac{1}{\zeta l J}\left(E_1^2 + E_2^2 +\zeta^2 E_3^2\right) - \frac{g l}{\zeta} \left(\zeta^2 B_1^2 +\zeta^2 B_2^2 + B_3^2\right), \label{L3consim} 
\end{align}
with $\zeta$ being the ration between the layer spacing and the lattice length $l$. The Lagrangian may be re-written in the form of \rf{L2D} with the metric
\begin{align}
 g_{3D} := \exp(2\Theta)\left[ \begin{array}{cccc}
                  - 1 & 0 & 0 & 0 \\
		  0 & \frac{1}{9Jgl^2}& 0 & 0 \\
		  0 & 0 & \frac{1}{9Jgl^2} & 0 \\
		  0 & 0 & 0 & \frac{1}{9Jgl^2\zeta^2}
                  \end{array}
\right] , \label{prelim3dm}
\end{align}
if the condition 
\begin{align}
 3 g = 16 J ,\label{metricitycond}
\end{align} 
which ensures consistency of the metrical interpretation, is satisfied. We obtain an arbitrary conformal factor $\exp(2\Theta)$ because in $3+1$ dimensions, the conformal factor of the metric in the action of Yang-Mills exactly simplifies and does not appear in the action. The last coordinate of the metric is in the direction of the stacking of the layer. Thus if we now stack shells of spherical layers together the effective metric will look like
\begin{align}
  g_{3D} := \exp(2\Theta) \left\{ -dt^2 + \frac{dr^2}{9Jgl^2 \zeta^2}+ \frac{r^2}{9 Jgl^2} d^2\Omega\right\} , \label{echtemetrik3d}
\end{align}
with $r$ being the radial distance and $d^2\Omega$ being the standard metric on the 2-sphere. Satisfying the metricity constraint \rf{metricitycond}, we automatically have that at cold enough temperatures, the system will be in the stringnet condensate phase of emergent light\cite{WenLab}. So we will have $U(1)$-Yang-Mills on the curved background. If we now wish that background to be Schwarzschild, we must have
\begin{eqnarray}
 \exp(2\Theta) & = 9 J g l^2 = \left(1-\frac{2M}{r}\right)c^2\nn
\zeta & = c \left(1-\frac{2M}{r}\right) \label{solschwarz} , 
\end{eqnarray}
where $M$ is the mass of the black hole and $c$ is the desired speed of light. Since $\zeta$,$g$,$J$, and $l$ must all be positive quantities, we see that we can build the black hole only down to the horizon, we cannot cross the horizon as this requires a sign change in the metric. We can thus build the Schwarzschild spacetime from infinity down to the horizon by appropriately tuning $g$ and $J$ (which here vary between $0$ and a value of order $1$ and is thus not problematic) and piling up the layers closer and closer together close to the horizon. We tune $g$ by changing the width of the superconducting film in which the spins lattices are embedded;  $J$ is tunes by changing the size of the holes inside the superconducting film which contain the spins.  For a small enough black hole, we can build the metric to within an arbitrarily small fraction of the effective Planck length of the horizon: to get to within $\epsilon$ of the horizon, the smallest ratio we need between the inter-layer distance and the intra-layer lattice distance is $\zeta= \frac{c \epsilon }{2M+\epsilon}$. As such, if quantum gravitational effects such as Hawking radiation are due simply to the existence of a horizon and if they exist even with a minimum length, these effects should be observable in this setup. 

On the other hand, if the interior of the black hole is also needed to reproduce the quantum gravitational effects\footnote{For example, if the entropy of a black hole really is entanglement entropy.}, then we also need to be able to build the interior of the black hole. For this we need a coordinate system where the signature of the metric does not jump. We can do this at the cost of having a time dependent metric for example with the Kruskal–Szekeres coordinates:  
\begin{align}
ds^2 = \frac{32 G^3 M^3}{r}e^{-\frac{r}{2GM}}\left(-dV^2+dU^2\right) + r^2 d^2\Omega \label{Krus}
\end{align}
where $V^2-U^2 = (1-\frac{r}{2GM})e^{\frac{r}{2GM}}$ or $\frac{r}{2GM} = 1 + W_0(\frac{U^2-V^2}{e})$ and $V^2 + U^2 = $. $W_0$ being the principle branch of the $W$ Lambert function. Therefore, replacing for $r$ we have the metric
\begin{align}
 g_{KS}= \frac{4(2GM)^2}{1+W_0(\frac{U^2-V^2}{e})}e^{-\left(1+W_0(\frac{U^2-V^2}{e})\right)}\left(-dV^2+dU^2\right) + \left(\frac{2GM}{1+W_0(\frac{U^2-V^2}{e})}\right)^2 d^2\Omega .\label{Kruskal}
\end{align}

It might be quite difficult to fit whole spacial slices of the Kruskal-Szekeres spacetime and so we will restrict ourselves to the equatorial plane and use 2-spacial dimension string-net condensate to build the equatorial plane. Because only one polarization of the emergent light is possible in the 2-spacial dimension string-net condensate we will also lose one of the polarization of the emergent light in the process. This however should have no effect on the possibility to observe quantum gravitational effects. We will also want to give ourselves more freedom by shaping the 2D spacial surface of sting-net condensate into a surface of revolution, a surface with polar symmetry. We may describe such a surface with a function with a function $\rho(r)$ which gives the radius of the circle at radial distance $r$, so that for a plane for example $\rho(r)=r$ and for a 2-sphere of radius $R$ we have $\rho(r) = R\sin(r/R)$. As such, the metric for such a surface of revolution is 
\begin{align}
 \gamma = -dt^2 + d^r + (\rho(t,r))^2 d\phi^2 , \label{metricsurf}
\end{align}
where we admitted the possibility that the surface might change its shape in time. Thus if we put the string-net condensate on such a curved surface its Lagrangian will be
\begin{align}
 \Lag_{\gamma} = -\sqrt{\det(\gamma)} \gamma^{ik} \gamma^{jm} \tilde{F}_{ij}(A)\tilde{F}_{km} , \label{pregam}
\end{align}
where 
\begin{align}
 \tilde{F_{0i}} & = \frac{F_{0i}}{\sqrt{J}} \nn
 \tilde{F_{ij}} & = \sqrt{g}l{F_{ij}} \ \ i,j\neq 0.  
\end{align}
The black hole metric in the equatorial plane can be written as 
\begin{align}
 g_{KS}= \frac{4(2GM)^2}{1+W_0(\frac{U^2-V^2}{e})}e^{-\left(1+W_0(\frac{U^2-V^2}{e})\right)}\left(-dV^2+dU^2\right) + \left(\frac{2GM}{1+W_0(\frac{U^2-V^2}{e})}\right)^2 d\phi^2 .\label{Kruskaleq}
\end{align}
We will identify $V$ with the time coordinate of the laboratory time $t$ and $U$ with the laboratory radial coordinate of the surface $r$. Thus in order to mimic the equatorial slice of the black hole in Kruskal-Szekeres coordinates we need to satisfy
\begin{align}
 \sqrt{-\det (g_{KS})}g_{KS}^{tt}g_{KS}^{ss} & = \frac{\exp(2\theta)}{J} \sqrt{-\det(\gamma)}\gamma^{tt}\gamma^{ss} \ \ s\in\{r,\phi\} \nn
 \sqrt{-\det(g_{KS})}g_{KS}^{rr}g_{KS}^{\phi \phi} & = gl^2\exp(2\theta) \sqrt{-\det(\gamma)}\gamma^{rr}\gamma^{\phi\phi} , \label{jackal}
\end{align}
where in the last two equations we must remember that $\det(g_{KS}) = $ ${g_{KS}}_{UU}{g_{KS}}_{VV}{g_{KS}}_{\theta \theta}|_{\theta=\pi/2}{g_{KS}}_{\phi \phi}|_{\theta=\pi/2}$, and we we still have a free conformal factor $\exp(2\theta)$ because even though we have restricted to a 2+1 dimensional subspace, the theory is still Yang-Mills in 3+1 dimensions. By multiplying the first and second line of \rf{jackal} we see that here too we have a metricity constraint:
\begin{align}
 1= \frac{gl^2 \exp(4\theta)}{J(\rho(t,r))^2} .\label{metricity2}
\end{align}

Putting everything together, we obtain the equatorial plane of the maximal analytical extension of the Schwarzschild black hole with emergent $U(1)$ Yang-Mills if we set the following values\footnote{It is interesting to compare the function $\rho$ here to the suggestion, arrived at totally differently, of \cite{graphene}}:
\begin{align}
 \exp(2\theta) & = J \rho \nn
Jgl^2 & = 1 \nn
\rho & = \frac{e^{\half\left(1+W_0(\frac{r^2-t^2}{e})\right)}}{4\sqrt{\left(1+W_0(\frac{r^2-t^2}{e})\right)}} . \label{result}
\end{align}

\section{Conclusion}
We have established here a prescription to investigate quantum fields in curved spacetime. The first step consists of finding a condensed matter system which has the desired quantum field as an effective quantum field in an appropriate limit. The second step is then to locally tune the coupling constants of the condensed matter system in order that the effective quantum field also sees the desired metric as an effective metric. 

Because the universe is very flat on the scales on which matter is described quantum mechanically, it is very difficult to experiment with quantum fields on curved spacetimes. Such modelling of analogue systems in the lab can provide valuable experimental opportunities and insight which cannot be gained otherwise. Of particular are kinematical quantum gravitational effects such a Hawking radiation. And the investigation of how much such effect depend on the continuum and how much they are affected by a ``fundamental'' discreteness scale. 

\section{Acknowledgments}
The author would like to thank Lorenzo Sindoni, Mercedes Mart\'{\i}n-Benito, Luis Garay, Sabine Hossenfelder, Carlo Rovelli and Louis Crane for discussions and suggestions. 
\appendix 
\section{When is variable speed curvature?}\label{speedmet}
What are the functions $f:= \tilde{k}^2$ which satisfy \rf{adlin}? Satisfying \rf{adlin} implies that $f$ is a squared norm comming from an innerproduct so we may write 
\begin{align}
f(v;x) = v^T M(x) v , \label{ap1}
\end{align}
where $M(x)$ is a position dependent $D\times D$ symmetric and positive-definite square matrix, $v$ is a D-dimensional vector and $x$ is a position coordinate in the D-dimensional flat space. We may locally orthnomarly diagonalize $M$. For example in D=2, we may write
\begin{align}
M(x) = \left[ \begin{array}{cc} \cos(\phi(x)) & -\sin(\phi(x)) \\ \sin(\phi(x)) & \cos(\phi(x)) \end{array} \right] \left[ \begin{array}{cc} a^2(x) & 0 \\ 0 & b^2(x) \end{array} \right] \left[ \begin{array}{cc} \cos(\phi(x)) & \sin(\phi(x)) \\ -\sin(\phi(x)) & \cos(\phi(x)) \end{array} \right] . 
\end{align}
Writing v as 
\begin{align}
v= \lambda \left[ \begin{array}{c}
\cos(\theta)\\
\sin(\theta)
          \end{array}\right] ,
\end{align}
we obtain
\begin{align}
f(\lambda,\theta;x) = \lambda^2 \{a^2(x) \cos^2(\theta-\phi(x)) + b^2(x) \sin^2(\theta-\phi(x))\} . 
\end{align}
The inverse of the speed squared must thus be given by an ellipse\footnote{Or in higher dimensions, a (hyper)-ellipsoid.} at every point:
\begin{align}
\frac{1}{c^2(\theta,x)} = a^2(x) \cos^2(\theta-\phi(x)) + b^2(x) \sin^2(\theta-\phi(x)), \label{spdfct}
\end{align}
where we describe the 2-dimensional unitary vector $u$ by its angle with respect to the x-axis. 
We see that the degrees of the ellipse ($a$, $b$ and $\phi$) are exactly the three degrees of freedom of a metric at every point. 
Hence, we now see that the speed function is highly contrived, one cannot have for example a ``flower'' as a speed function:
\begin{align}
c(\theta) = 1+ \half \cos(N\theta) , \ \ N\in\N .
\end{align}
Nevertheless, naturally arising systems will satisfy \rf{spdfct}.

\bibliographystyle{iopart-num}
\bibliography{bib}
\end{document}